\documentclass[a4paper]{jpconf}
\usepackage{graphicx}

\usepackage{ amssymb }
\graphicspath{{plots/}}

\begin{document}
\title{Results from MINOS and NO$\nu$A}

\author{C Backhouse}

\address{California Institute of Technology\\M/C 356-48, 1200 E California Blvd, Pasadena, CA 91125, USA}

\ead{bckhouse@caltech.edu}

\begin{abstract}

The MINOS experiment, operating in the NuMI beam since 2005, has provided the
most precise measurement of the atmospheric mass splitting $|\Delta m^2_{32}|$,
and the recent combination of the $\nu_\mu$, $\nu_e$, and atmospheric neutrino
samples has provided some evidence of non-maximal mixing, and hints about the
neutrino mass hierarchy and the $\theta_{23}$ octant. Construction of the
NO$\nu$A experiment, situated off-axis in the upgraded NuMI beam, is almost
complete. Over the coming years it will have significant power to probe the
questions of the mass hierarchy, $\theta_{23}$ octant, and the possibility of
$\mathcal{CP}$ violation in the lepton sector. This paper gives an overview of
the results from MINOS, and of the sensitivity of the NO$\nu$A experiment.
\end{abstract}

\section{Introduction}

The MINOS and NO$\nu$A experiments are both situated on the NuMI neutrino
beam-line. 120GeV protons from the Fermilab Main Injector strike a graphite
target, the resulting pions are focussed by two magnetic horns, and decay to
produce a beam of primarily muon neutrinos. Observing the energy-dependent
disappearance of these muon neutrinos, and subdominant appearance of electron
neutrinos provides information about the neutrino oscillation parameters. Both
experiments make use of a two-detector design, in which observations by a Near
Detector allow measurements of the beam flux, cross-sections and detector
effects, while a larger, but otherwise similarly constructed, Far Detector
observes the change in the beam composition over a much longer baseline.

\section{MINOS}

The MINOS detectors are magnetized steel-scintillator tracking sampling
calorimeters. They consist of alternating planes of 2.54cm thick steel and
$4.1$cm$\times1$cm solid scintillator strips. Each scintillator strip contains
an embedded wavelength-shifting fibre, read out at both ends by Hamamatsu
multi-pixel PMTs. Adjacent scintillator planes are oriented at $90^\circ$ to
each other, allowing for 3D event reconstruction. The detectors are magnetized
by a coil to a field of around 1T, allowing for charge identification of
muons from curvature.

The MINOS Far Detector is located underground in the Soudan mine, Minnesota,
735km from the NuMI target. It has a total mass of 5.4kton, and has been
operating since 2003. The Near Detector is located underground at Fermilab,
about 1km downstream from the target. It has a mass of about 1kton, and was
completed in 2005. Since 2005, the NuMI beam-line has delivered
$10.7\times10^{20}$ protons-on-target (POT) to MINOS in the ``low energy''
neutrino configuration, and $3.36\times10^{20}$ POT in antineutrino
configuration, plus smaller samples in special configurations. In addition,
MINOS has collected an exposure of 37.8 kton years of atmospheric neutrinos. In
2013, following a shutdown, the NuMI beam returned, delivering neutrinos to
NO$\nu$A (and MINOS+ and MINER$\nu$A) in the new ``medium energy''
configuration, optimized to give the greatest flux around oscillation maximum
($\sim2$GeV) given NO$\nu$A's off-axis location.

\begin{figure}
  \begin{minipage}{.65\linewidth}
    \includegraphics[width=.49\linewidth]{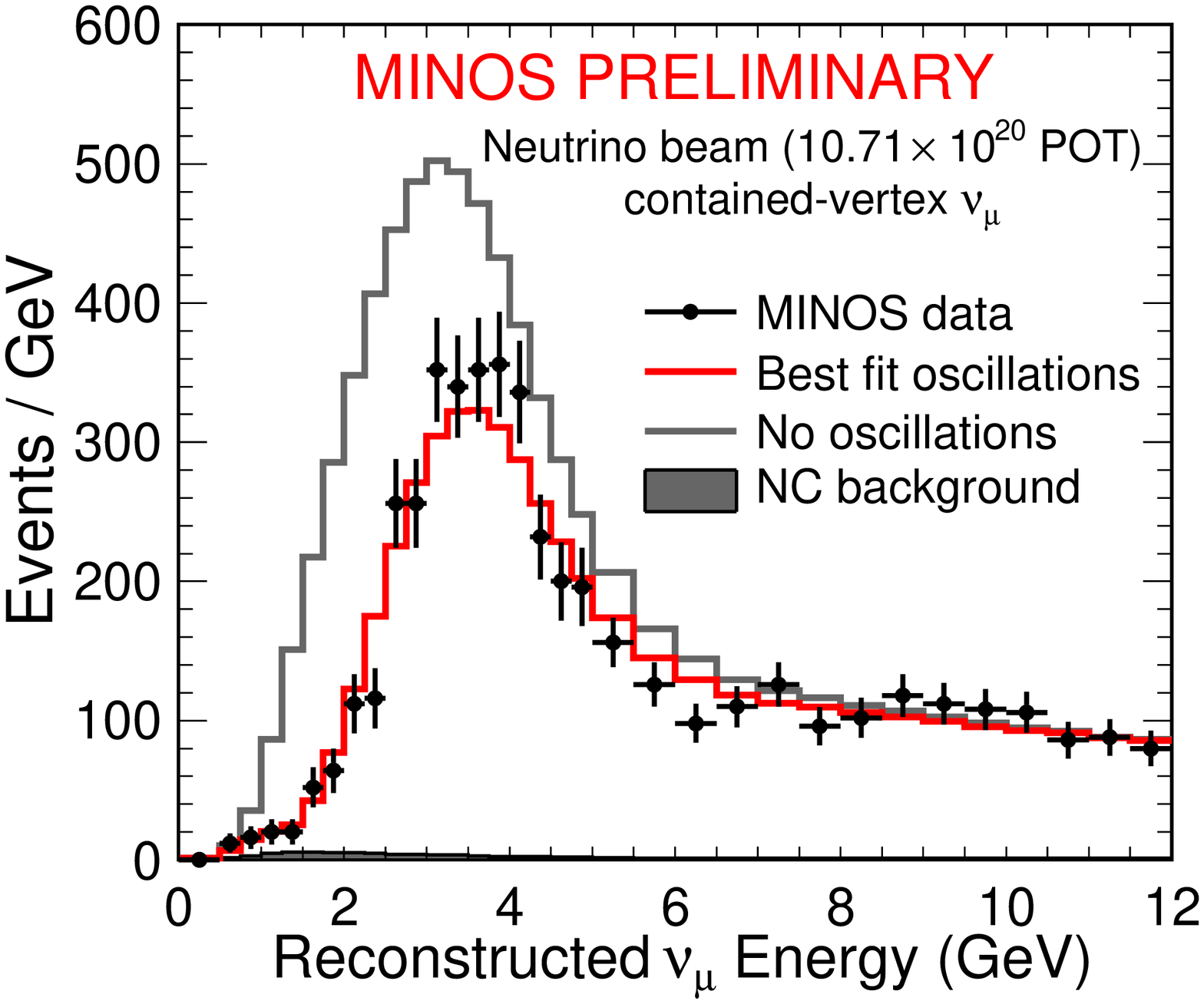}
    \includegraphics[width=.49\linewidth]{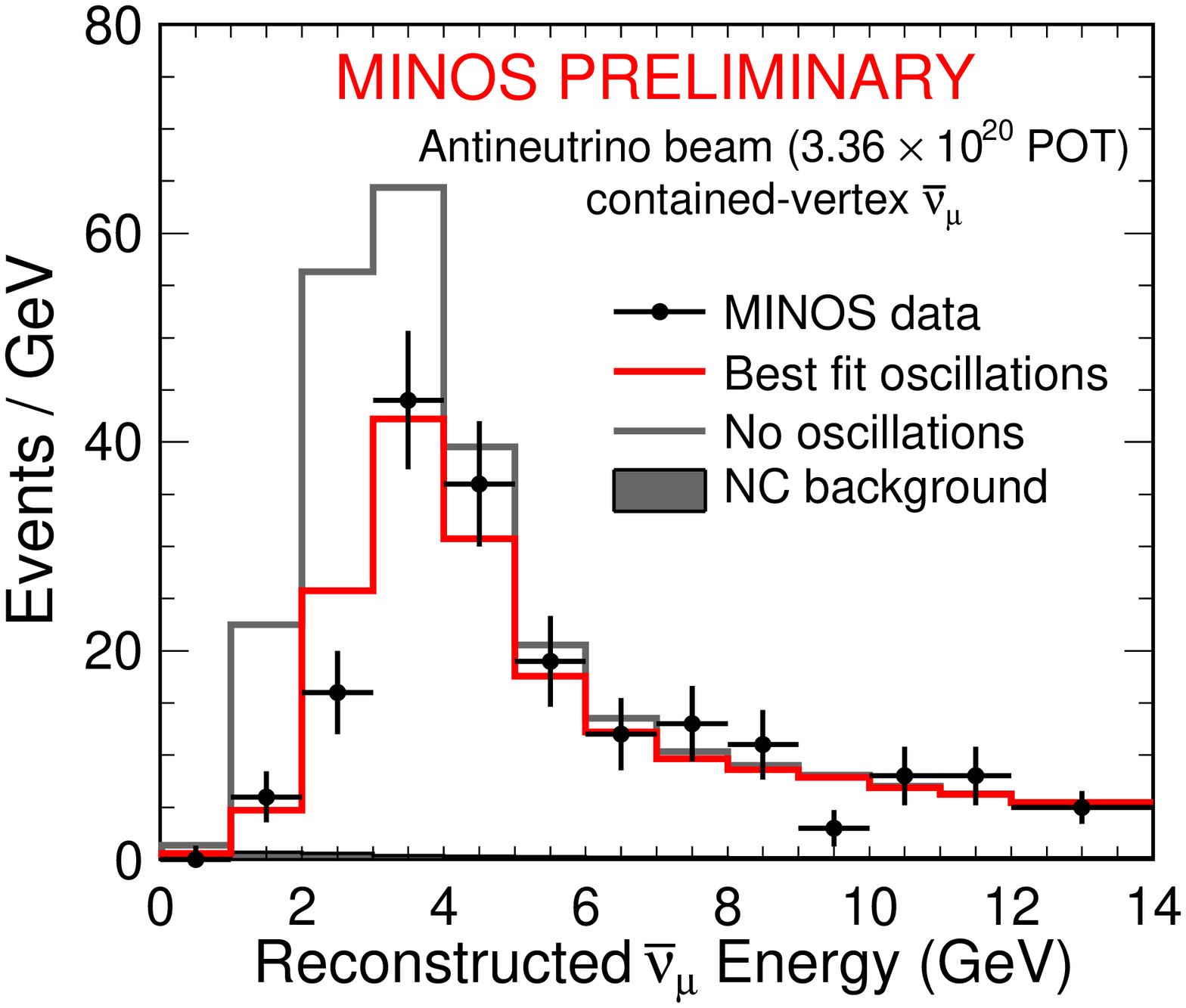}
    \vspace{-1.5em}\\
    \includegraphics[width=.49\linewidth]{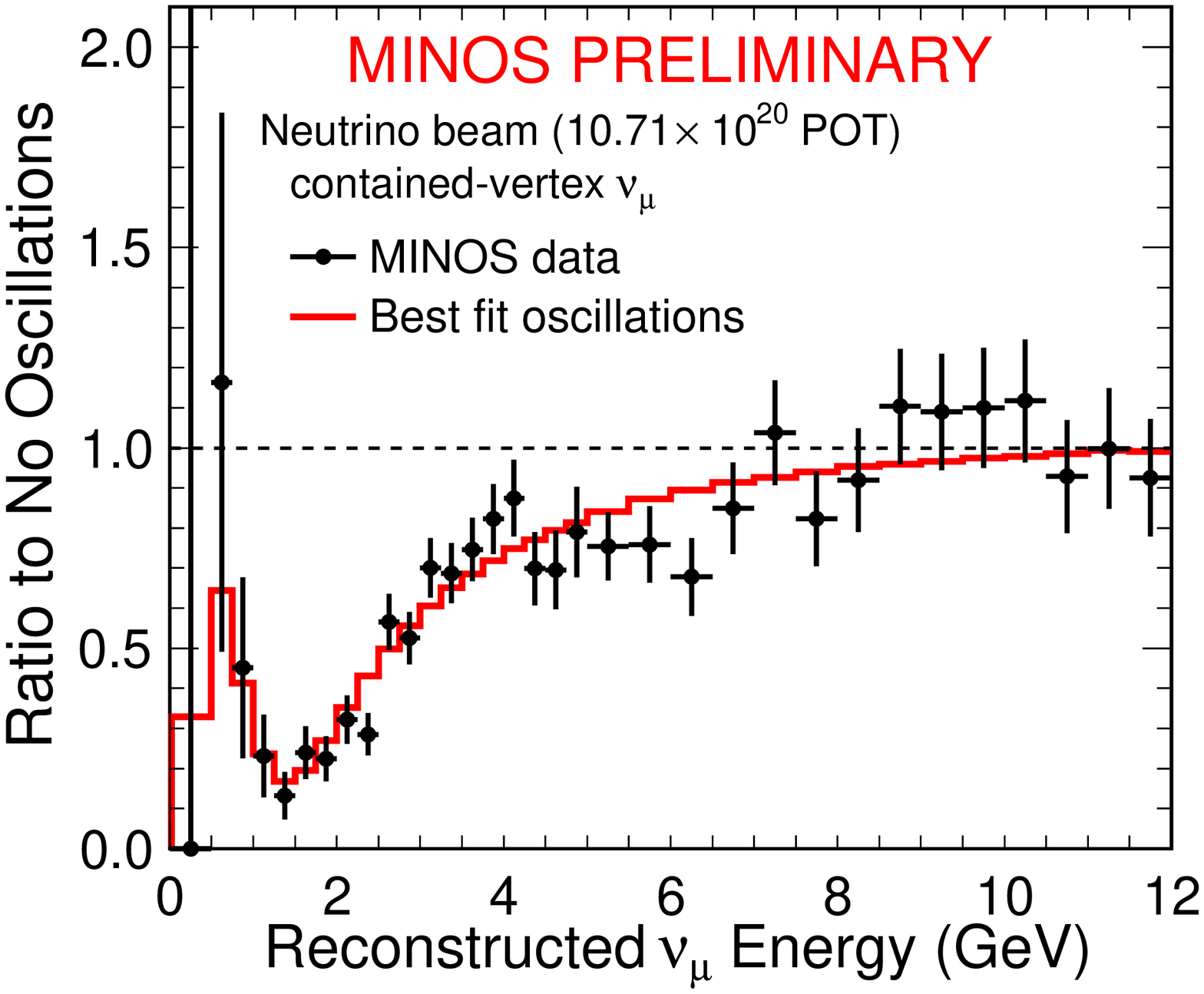}
    \includegraphics[width=.49\linewidth]{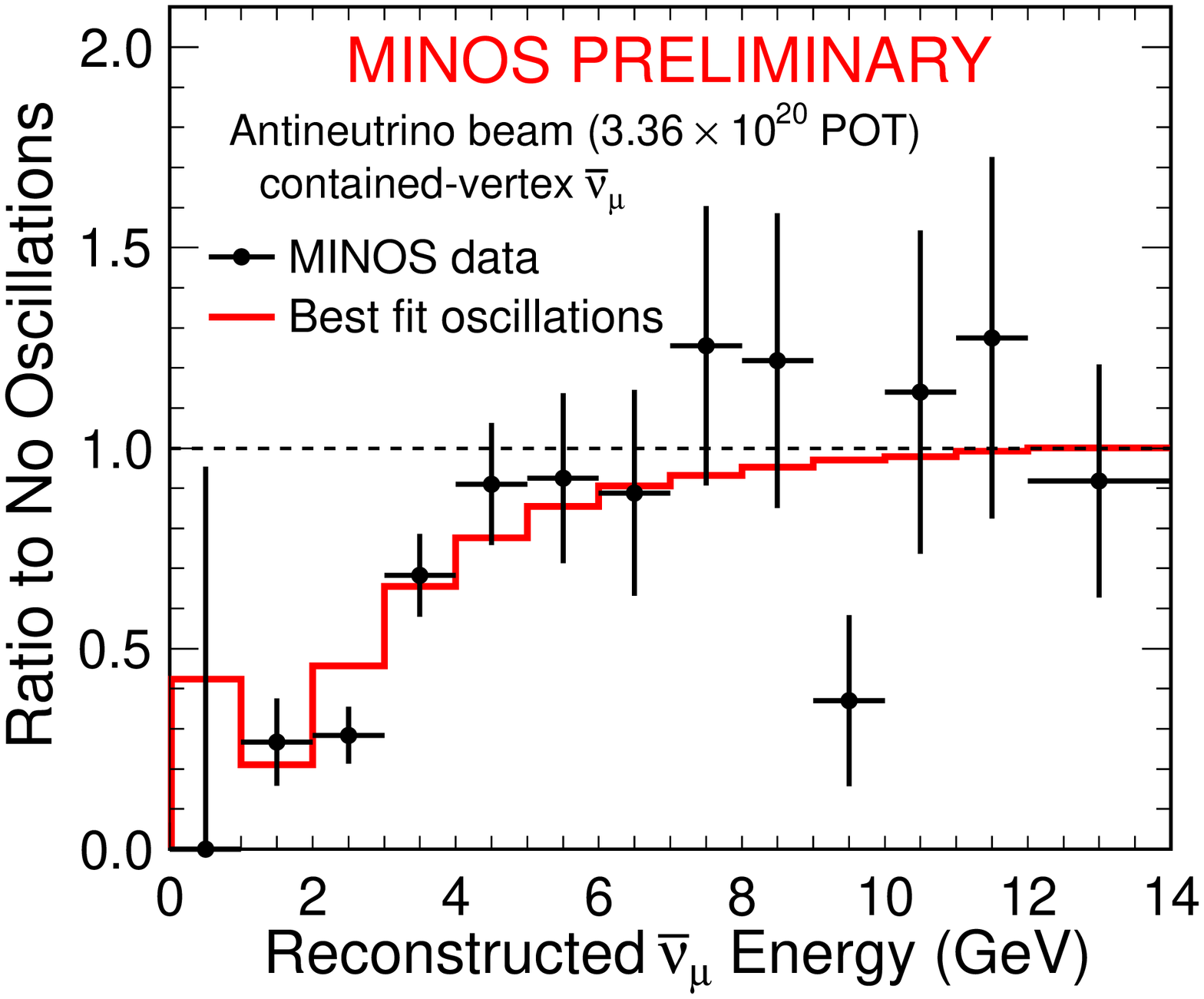}
  \end{minipage}
  \hspace{.025\linewidth}
  \begin{minipage}{.3\linewidth}
    \caption{Energy spectra from the MINOS neutrino disappearance analyses. The
      black points are the observed event counts, with statistical errors. The
      grey curves show the expectation in the absence of oscillations, and the
      red curves show the best oscillation fit. Agreement with the oscillation
      hypothesis is very good for both neutrino mode (left) and antineutrino
      mode (right). The lower panels display ratios to the unoscillated
      prediction.}
    \label{fig:beam_spectra}
  \end{minipage}
\end{figure}

Extrapolating from the observed neutrino spectrum in the Near Detector, in the
absence of oscillations, MINOS would expect to select 3201 $\nu_\mu$
interactions in the Far Detector, and 363 interactions in the antineutrino
running. The observed counts are 2579 and 312 respectively. Figure
\ref{fig:beam_spectra} shows the spectral shapes of these deficits, compared to
the best fit from the neutrino oscillation hypothesis. The three-flavour
oscillation paradigm fits the data well: 18\% of Monte Carlo pseudo-experiments
have a worse fit \cite{ref:minos_numu}.

\begin{figure}
  \begin{centering}
    \includegraphics[width=.9\linewidth]{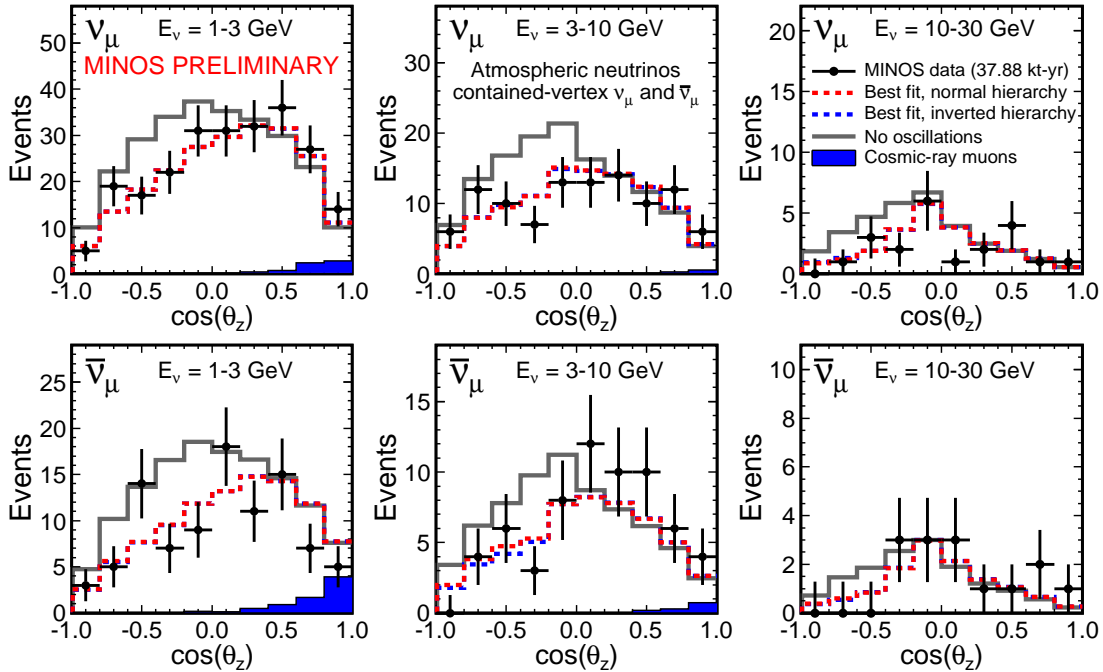}\\
  \end{centering}
  \caption{Spectra for the MINOS atmospheric neutrino sample. The points show
    the observed event counts as a function of reconstructed zenith angle. The
    grey curves show the unoscillated predictions, and the red and blue dashed
    curves are the best oscillation fits assuming normal or inverted hierarchy
    respectively. The filled blue histograms indicate the background due to
    cosmic-ray muons.
  }
  \label{fig:atmo_spectra}
\end{figure}

The inclusion of atmospheric neutrino samples into the analysis provides
additional sensitivity, particularly to the mass hierarchy. The unoscillated
expectation is for 1100 events, 905 are observed. Figure~\ref{fig:atmo_spectra}
shows the spectra of the contained muon-selected events as a function of zenith
angle. Partially-contained and showering events are also included in the
analysis as additional samples.

\begin{figure}
  \begin{minipage}{.65\linewidth}
    \includegraphics[clip,width=\linewidth]{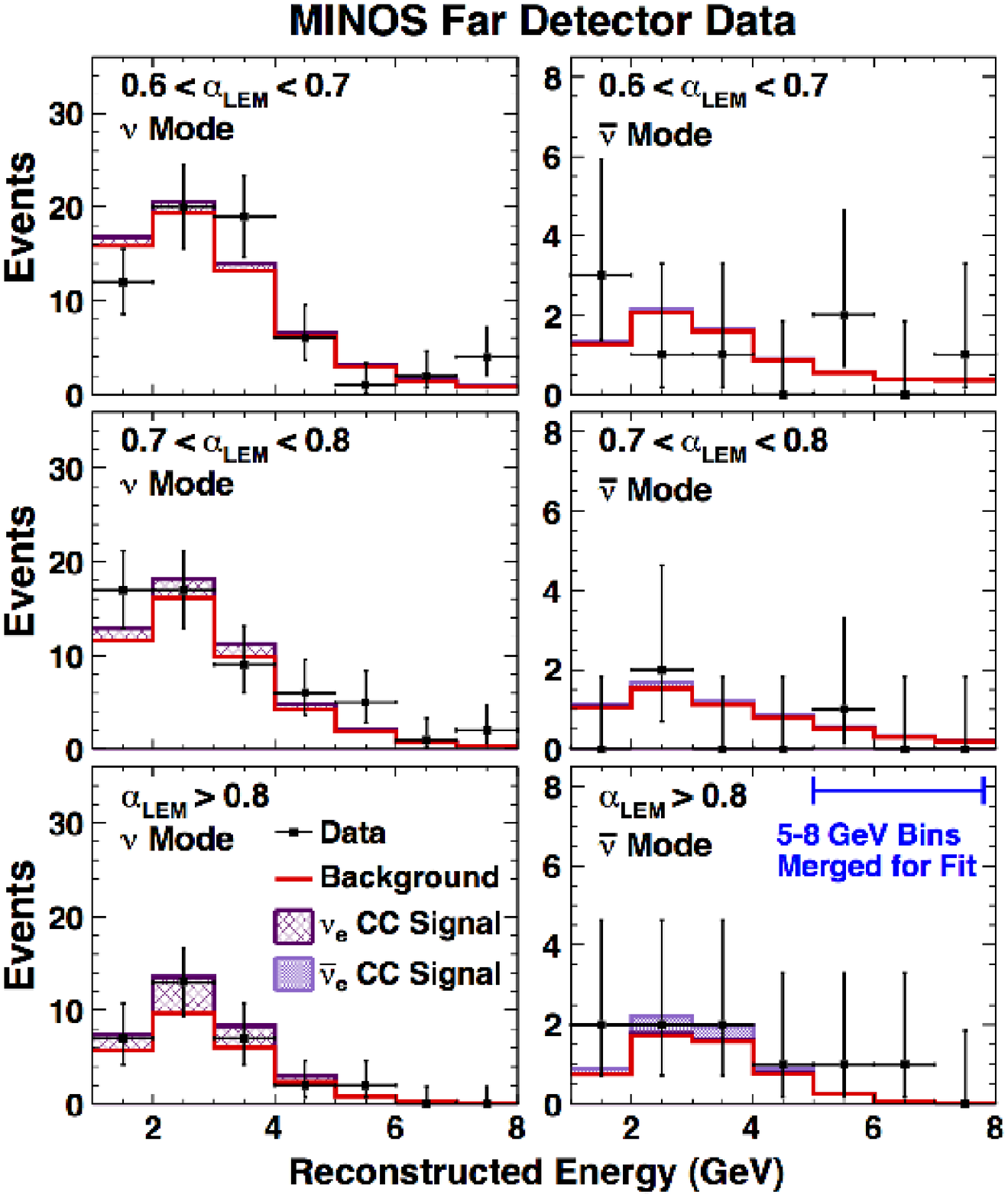}
  \end{minipage}\hspace{.025\linewidth}
  \begin{minipage}{.345\linewidth}
    \caption{Spectra of $\nu_e$-selected events in MINOS. The points are the
      number of events selected as a function of reconstructed neutrino
      energy. The red curve shows the prediction for $\theta_{13}=0$, and the
      purple region shows the excess for the best-fit value of $\theta_{13}$.
      The samples are divided by PID value (top to bottom) and between neutrino
      (left) and antineutrino (right) running. The data is divided into bins of
      PID value, from the lowest selected PIDs (top) to the most $\nu_e$-like
      events (bottom). 88 neutrino-mode events and 12 antineutrino-mode events
      are selected, compared to an expectation for $\theta_{13}=0$ of 69.1 and
      10.5. The expectations for $\theta_{13}=0.1$ are 95.1 and 13.6.
    }
    \label{fig:nue_spectra}
  \end{minipage}\\
\end{figure}

Selecting electron neutrino candidates, the expectation for $\theta_{13}=0$ is
69.1 events of background in neutrino mode, and 10.5 in antineutrinos. The
counts observed in data are 88 and 12. Figure \ref{fig:nue_spectra} shows these
excesses in bins of energy and PID value.
The hypothesis that $\theta_{13}=0$ is rejected at the 96\% confidence level
\cite{ref:minos_nue}.

\begin{figure}
  \begin{centering}
    \includegraphics[width=.9\linewidth]{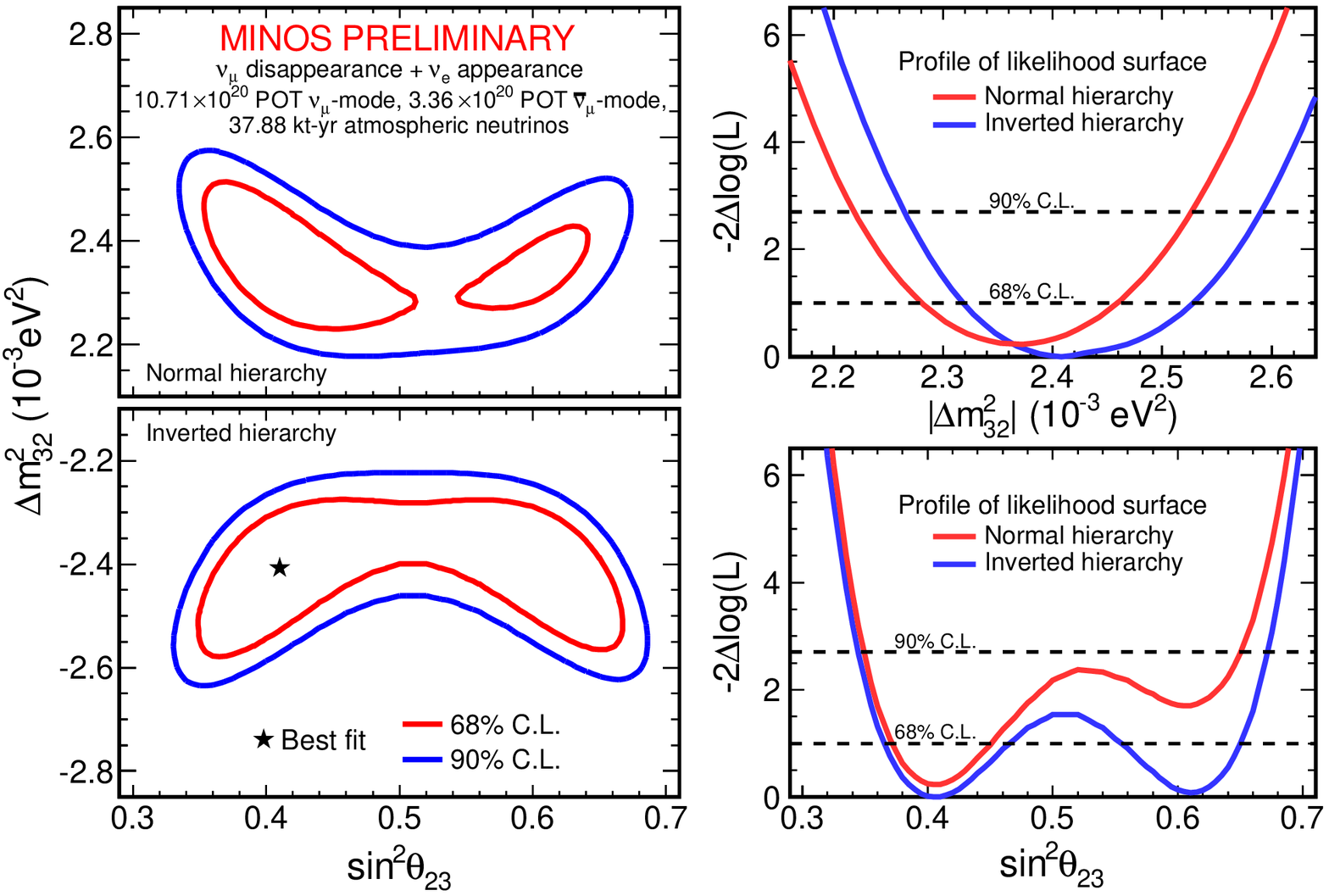}\\
  \end{centering}
  \caption{Combined results from the MINOS $\nu_\mu$, $\nu_e$, and atmospheric
    analyses. The star shows the best fit values of $\sin^2\theta_{23}$ and
    $\Delta m^2_{32}$. This point weakly favours non-maximal mixing, the lower
    octant, and inverted hierarchy. The red and blue contours indicate 68\% and
    90\% confidence levels. The panels on the right display the profile
    likelihood projection onto each of the two parameters.}
  \label{fig:minos_combined}
\end{figure}

\begin{figure}
  \begin{minipage}{.375\linewidth}
    \includegraphics[width=\linewidth]{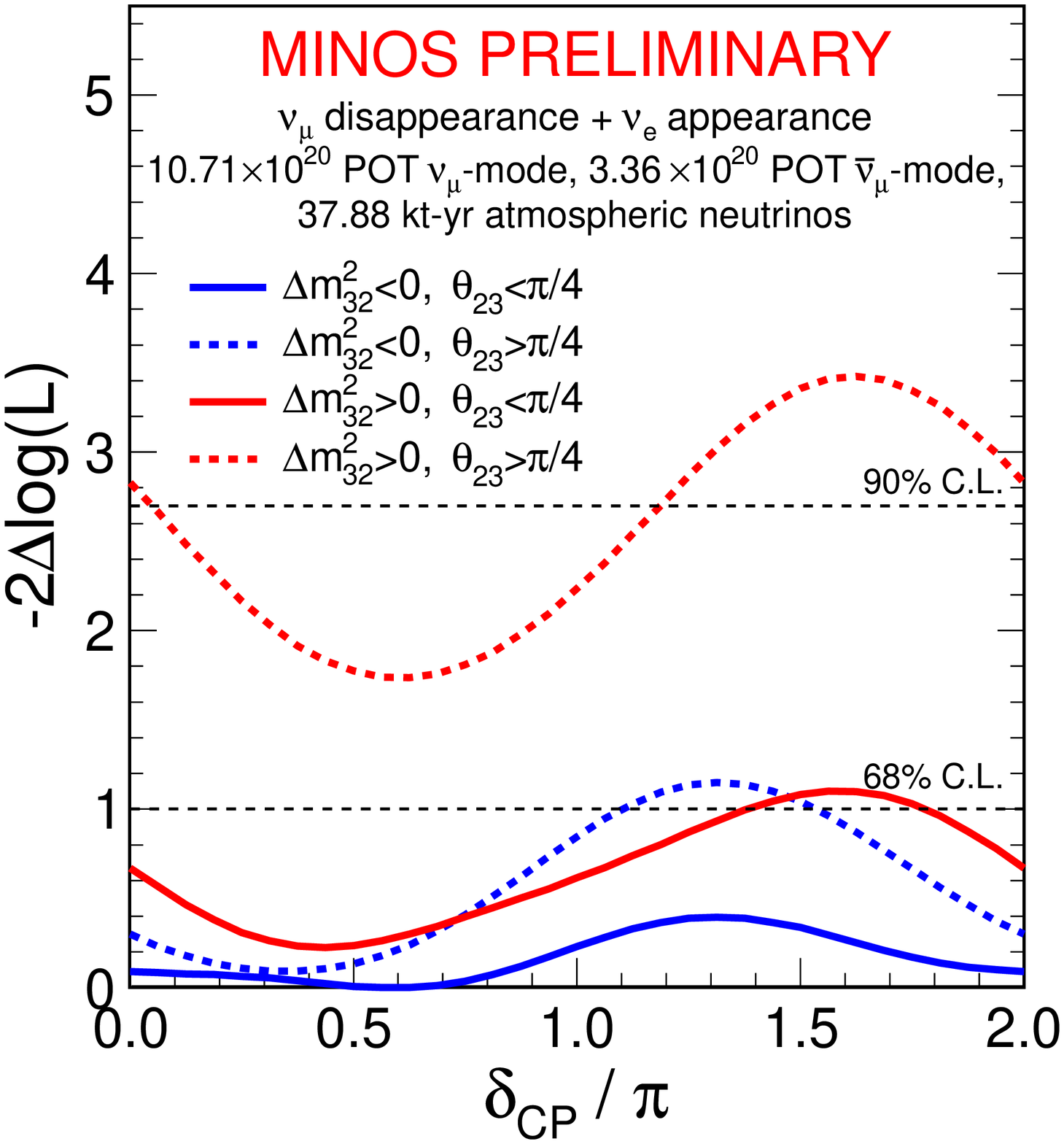}
  \end{minipage}\hspace{.025\linewidth}
  \begin{minipage}{.575\linewidth}
    \includegraphics[width=\linewidth]{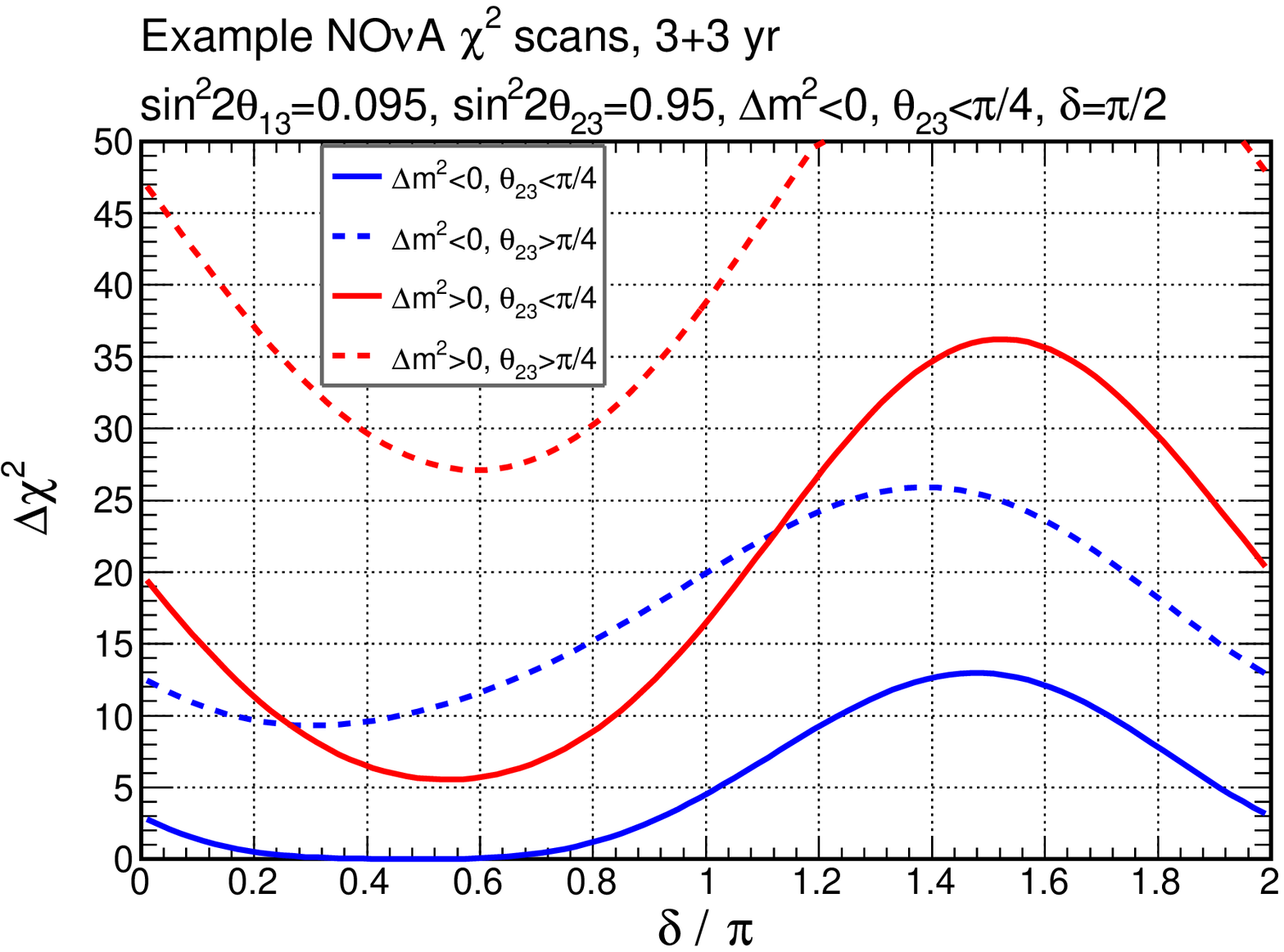}
  \end{minipage}\\
  \begin{minipage}{.375\linewidth}
    \caption{Log-likelihood differences from the best fit oscillation
      parameters for different values of $\delta_{CP}$, hierarchy and
      $\theta_{23}$ octant for the combination of MINOS analyses. The
      region normal hierarchy, upper octant,
      $\delta_{CP}\sim{3\pi\over2}$ is disfavoured at greater than 90\%
      confidence.}
    \label{fig:minos_slices}
  \end{minipage}\hspace{.025\linewidth}
  \begin{minipage}{.575\linewidth}
    \caption{Sensitivity of NO$\nu$A for a nominal run of three years neutrino
      plus three antineutrino ($18\times10^{20}$ POT each). The true
      oscillation values are chosen to be close to the MINOS best-fit. The
      shape of the curves is very similar to the MINOS results in
      Figure~\ref{fig:minos_slices}, but the expected sensitivity is much
      higher. For example, the most disfavoured scenario, normal hierarchy,
      upper octant, would be excluded at $5\sigma$ for all values of
      $\delta_{CP}$.}
    \label{fig:nova_slices}
  \end{minipage}
\end{figure}

Combining the information provided by the $\nu_\mu$, atmospheric, and $\nu_e$
samples, plus additional constraints on $\theta_{13}$ from reactor experiments
provides some sensitivity to the neutrino hierarchy and $\theta_{23}$ octant.
The solar parameters ($\Delta m^2_{21}$, $\theta_{12}$) are held fixed in this
analysis. $\theta_{13}$ is treated as a nuisance parameter, constrained by
external reactor results. $\delta_{CP}$, $\theta_{23}$, and $\Delta m^2_{32}$
are unconstrained and determined by the joint best fit to all the
spectra. Major sources of systematic error are included as nuisance parameters.

Figure \ref{fig:minos_combined} shows the confidence contours obtained in
($\Delta m^2_{32}$, $\sin^2\theta_{23}$) space. The best fit value of
$\sin^2\theta_{23}$ is 0.41, and maximal mixing is disfavoured at 79\%
confidence. The data also weakly favour the inverted hierarchy and the lower
$\theta_{23}$ octant. Figure \ref{fig:minos_slices} shows a different
projection of this information, with $\delta_{CP}$ information included. The
most disfavoured scenario is the combination of normal hierarchy and upper
$\theta_{23}$ octant which is rejected at the $81\%$ confidence level for all
values of $\delta_{CP}$. This measurement is also the most precise
determination of $|\Delta m^2_{32}|$. The results are summarized numerically in
Table \ref{tbl:summary} \cite{ref:minos_comb}.

\begin{table}
  \caption{Summary of MINOS combined oscillation fit results for normal and inverted hierarchy.}
  \label{tbl:summary}

  \begin{centering}
    \begin{tabular}{lrlrl}
      \br
      Parameter & Best fit & Confidence limits (NH) & Best fit & Confidence limits (IH) \\
      \mr
      $|\Delta m^2_{32}|/10^{-3}{\rm eV}^2$ & 2.37 & 2.28 - 2.46 (68\% C.L.) & 2.41 & 2.32 - 2.53 (68\% C.L.)\\
      $\sin^2\theta_{23}$ & 0.41 & 0.35 - 0.65 (90\% C.L.) & 0.41 & 0.34 - 0.67 (90\% C.L.)\\
      \br
    \end{tabular}\\
    \begin{tabular}{rl}
      Preference for inverted hierarchy:&$-2\Delta\log L$=0.23\\
      Preference for lower octant:&$-2\Delta\log L$=0.09\\
      Exclusion of maximal mixing:&$-2\Delta\log L$=1.54 ($\to 79\%$ C.L.)
    \end{tabular}\\
  \end{centering}
\end{table}

If some component of the disappearing MINOS flux were oscillating to sterile
neutrinos, one would expect to also see a deficit of neutral current events, in
addition to the deficit seen in $\nu_\mu$ charged currents. In fact, a search
for this effect found a small excess of neutral current interactions at the Far
Detector as compared to predictions. Interpreting in terms of oscillations
around $\Delta m^2\sim0.5{\rm eV}^2$
leads to a limit $\sin^22\theta_{\mu e} < 7.1\times10^{-3}$ at the 90\%
confidence level.

The MINOS detectors continue to operate in the NO$\nu$A-era beam. The higher
energy peak of the neutrino flux reduces the statistics acquired in the
oscillation region
, but the statistics at higher energies are greatly increased. MINOS+ expects
to collect around 4,000 $\nu_\mu$ charged current events per year. This
high-statistics sample will allow tests of the standard three-flavour
oscillation paradigm, including searches for sterile neutrinos, and sensitivity
to signatures of exotic physics such as large extra dimensions.

\section{NO$\nu$A}

The NO$\nu$A detectors are fine-grained, low-Z, highly active tracking
calorimeters. Cells of extruded PVC, approximately 6cm$\times$4cm are
formed into planes, with the orientation of successive planes alternating as in
MINOS. They are filled with a mineral oil/liquid scintillator mix, which
comprises 64\% of the detector by mass. Each cell contains a looped
wavelength-shifting fibre, read out at one end by one of the 32 pixels of a
Hamamatsu avalanche photodiode (APD), cooled to $-15^\circ{\rm C}$ \cite{ref:tdr}.

The Far Detector is located on the surface at Ash River, Minnesota, 810km from
the NuMI target. The detector consists of 344,000 channels, approximately
$16\times16\times60$m, for a total mass of 14kton. Assembly and placement of
the plastic blocks, scintillator filling, and electronics installation proceed
in parallel. All blocks are now in place, and oil filling is almost
complete. Over 2/3 of the detector is instrumented with APDs and being read
out. The Near Detector is located underground at Fermilab, approximately 1km
from the target. It comprises 18,000 channels, with a mass of 0.3kton. All
blocks are in place, and oil filling and electronics outfitting are in
progress.
The NuMI beam is being upgraded from a maximum capacity of 350kW up to
700kW. This will be achieved by converting the Recycler from running
antiprotons to protons, and shortening the cycle in the Main Injector from 2.2s
to 1.33s. The beam has been running stably since returning from shutdown in
September 2013, and power is gradually being increased. The maximum power
achievable will be limited to 500kW until upgrading of the Booster RF cavities
is complete. Thus far, in excess of $1.5\times10^{20}$ protons have been
delivered.

\begin{figure}
  \begin{minipage}{.475\linewidth}
    \includegraphics[width=\linewidth]{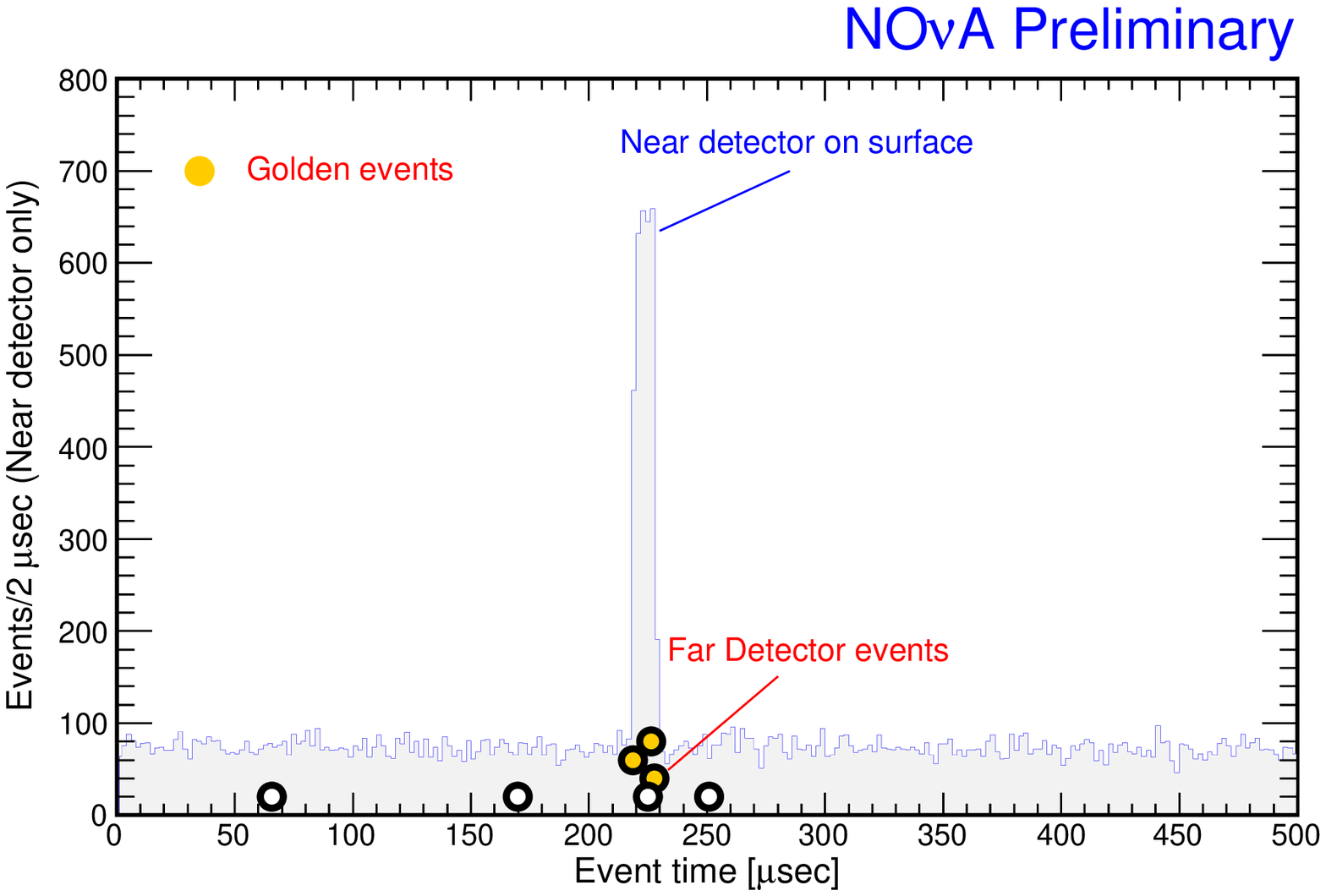}
  \end{minipage}
  \hspace{.025\linewidth}
  \begin{minipage}{.5\linewidth}
    \caption{The timing peak formed by neutrinos from the NuMI beam in the NDOS
      detector (filled histogram) with the times of selected interactions in
      the NO$\nu$A Far Detector marked (circles). The Far Detector events were
      found by a combination of hand-scanning and automated analysis.
      The estimated cosmic-ray background in the 10$\mu$s Far Detector spill
      window is 0.05 events.}
    \label{fig:numi_peak}
  \end{minipage}
\end{figure}
\begin{figure}
  \begin{minipage}{.45\linewidth}
    \caption{Event display of the first neutrino candidate event selected in
      the NO$\nu$A Far Detector. Each square represents energy deposited and
      detected in a single cell. The neutrino beam enters from the left. The
      top panel shows the information from vertically-oriented cells, with
      horizontal cells shown in the bottom panel. The overlaid coloured lines
      are the reconstructed paths of particles, and the cross marks the
      reconstructed position of the interaction vertex.}
    \label{fig:evd}
  \end{minipage}
  \hspace{.025\linewidth}
  \begin{minipage}{.475\linewidth}
    \includegraphics[width=\linewidth]{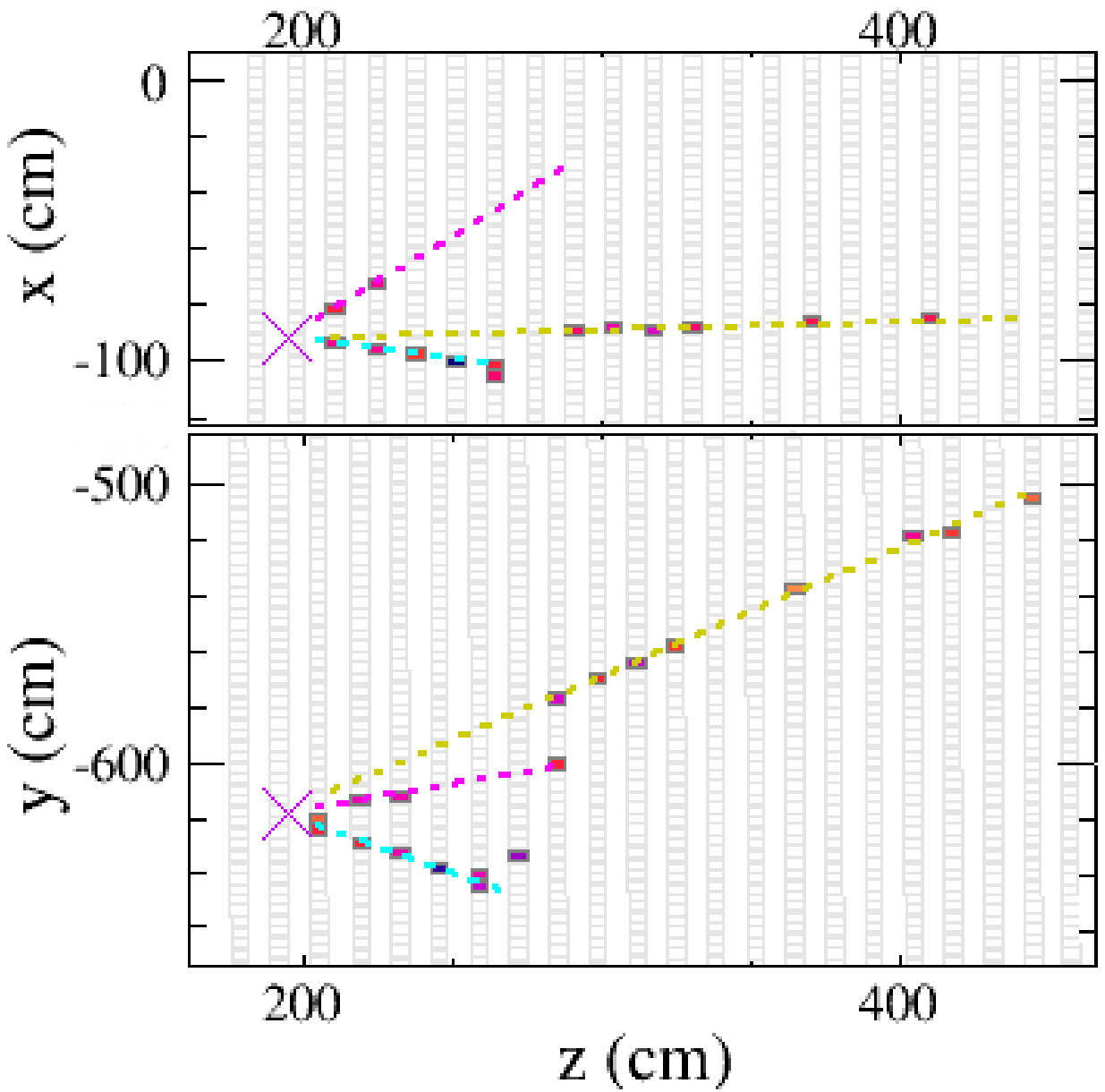}
  \end{minipage}
\end{figure}

It is important to validate the detectors' ability to pick neutrino
interactions out of the cosmic-ray background, and to verify the
synchronization of the Far Detector to the NuMI spill times. Analysis of the
Near Detector prototype on the surface at Fermilab (NDOS) shows a clear peak in
the neutrino candidate timing spectrum over the cosmic-ray
background. Propagating this timing information to the Far Detector gives a
10$\mu$s window in which neutrinos are expected to be found. A combination of
automated analysis and hand-scanning of events in a 500$\mu$s window around
this point (plus out-of-time spills for background estimation) found four
candidate events within the timing window, on a background of 0.05. The times
of the events relative to the beam window were kept secret until after the
choice of these events had been finalized. Figure \ref{fig:numi_peak} shows the
NDOS timing peak compared to the times of the selected neutrino candidates, and
Figure \ref{fig:evd} shows one of the Far Detector events in the timing window,
a multi-pronged interaction, with reconstructed neutrino direction coming from
Fermilab.

The sensitivity projections presented below assume $18\times10^{20}$ POT of
data taken in neutrino mode plus $18\times10^{20}$ POT taken with
antineutrinos. This corresponds to a nominal $6\times10^{20}$ POT per year and
a run plan with three years in each mode. The choice of what exposure to take
in each beam configuration is flexible, and may be adjusted in response to
changing circumstances. The oscillation parameters assumed are
$\sin^22\theta_{13}=0.095$
$\sin^22\theta_{23}=0.95$ or $1.0$ depending on context. We take $\Delta
m^2_{32}=2.35\times10^{-3}{\rm eV}^2$, and the other parameters from the latest
Particle Data Group averages \cite{ref:pdg}.

\begin{table}[h]
  \caption{Representative event counts for NO$\nu$A analyses. The columns
    represent $18\times10^{20}$ POT neutrino-mode and antineutrino-mode
    running.
    The $\nu_\mu$ analysis divides the dataset into three
    samples, candidate quasielastic events, non-quasielastic events, and events
    where the muon is not fully contained within the detector.
    The counts are constrained to the range 0-5GeV, the region of sensitivity
    to oscillations driven by the atmospheric mass splitting.}
\begin{centering}
\begin{minipage}{.35\linewidth}
  \begin{centering}
  \begin{tabular}{rrr}
    \br
    {\bf $\bf{\nu_e}$ selected} & $\nu$ & $\bar\nu$\\
    \mr
    NC & 19 & 10\\
    $\nu_\mu$ CC & 5 & $<1$\\
    Beam $\nu_e$ & 8 & 5\\
    \mr
    Tot bkg & 32 & 15\\
    Signal & 68 & 32\\
    \br
  \end{tabular}\\
  \end{centering}
\end{minipage}\hspace{.025\linewidth}%

\begin{minipage}{.6\linewidth}
  \begin{centering}
  \begin{tabular}{rrrrr}
    \br
    {\bf $\bf{\nu_\mu}$ selected} & $\nu_\mu$ CC & NC & $\bar\nu_\mu$ CC & NC\\
    \mr
    Quasielastic & 82 & $<1$ & 49 & $<1$\\
    non-QE & 168 & 14 & 78 & 6\\
    Uncontained & 233 & 6 & 134 & 3\\
    \br
  \end{tabular}\\
  \begin{centering}
    (0-5GeV visible energy)\\
  \end{centering}
  \end{centering}
  \vphantom{\ \\\ \\\ \\\ \\}
  \label{tbl:counts}
\end{minipage}\\
\end{centering}
\end{table}

\begin{figure}
  \begin{minipage}{.45\linewidth}
    \caption{Projected 90\% confidence contours in the ($\Delta m^2_{32}$,
      $\sin^2\theta_{23}$) space for two, six, or ten years of running
      (assuming $6\times10^{20}$ POT per year), for two values of
      $\theta_{23}$. In the case of $\sin^22\theta_{23}=0.95$, maximal mixing
      can be excluded at 90\% confidence in under two years of running.}
    \label{fig:nova_cc_conts}
  \end{minipage}
  \hspace{.025\linewidth}
  \begin{minipage}{.5\linewidth}
    \vspace{-2.5em}
    \includegraphics[width=\linewidth]{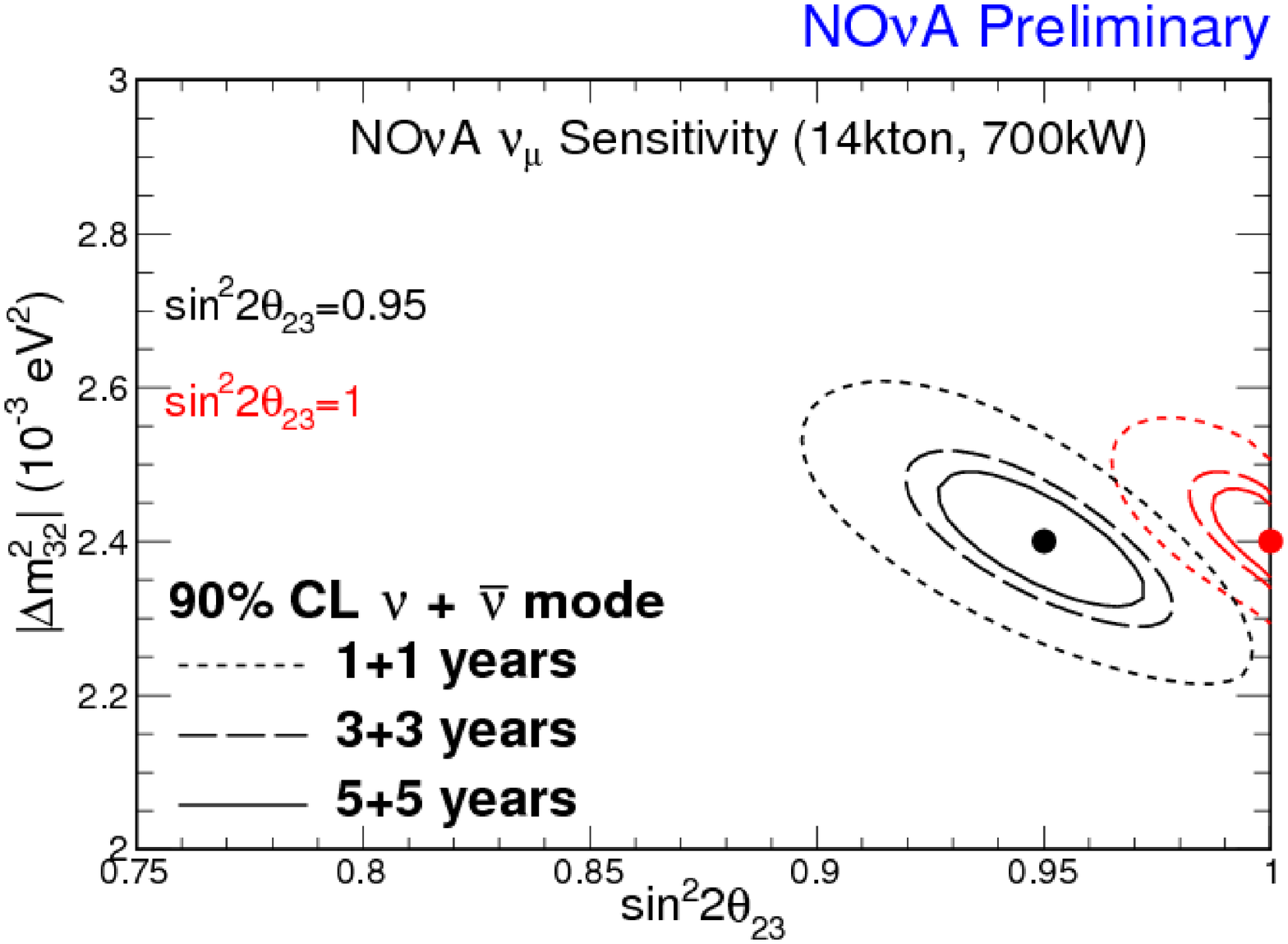}
  \end{minipage}
\end{figure}

Current global constraints on the atmospheric mixing angle are around
$\sin^22\theta_{23}\gtrsim0.9$, with hints of non-maximal mixing from
MINOS. The NO$\nu$A experiment will have significantly enhanced power to
distinguish between maximal and non-maximal mixing. The $\nu_\mu$ charged
current analysis being developed separates the data into three samples:
quasi-elastic candidates, which are very pure with good energy reconstruction;
the remainder of contained candidates; and uncontained events, where the muon
exits through the back or side of the detector, a sample with substantial
statistics, but poor energy resolution. Predicted event counts are given in
Table \ref{tbl:counts}.
The contours that would be obtained with increasing exposure are shown in
Figure \ref{fig:nova_cc_conts}, for two possible values of $\theta_{23}$. With
three years of neutrino running, and three years of antineutrinos, this
analysis should obtain percent-level uncertainty on the atmospheric mixing
parameters. If $\sin^2\theta_{23}=0.95$, maximal mixing could be excluded at
90\% confidence with one year of running.

\begin{figure}
  \begin{minipage}{.5\linewidth}
    \includegraphics[width=\linewidth]{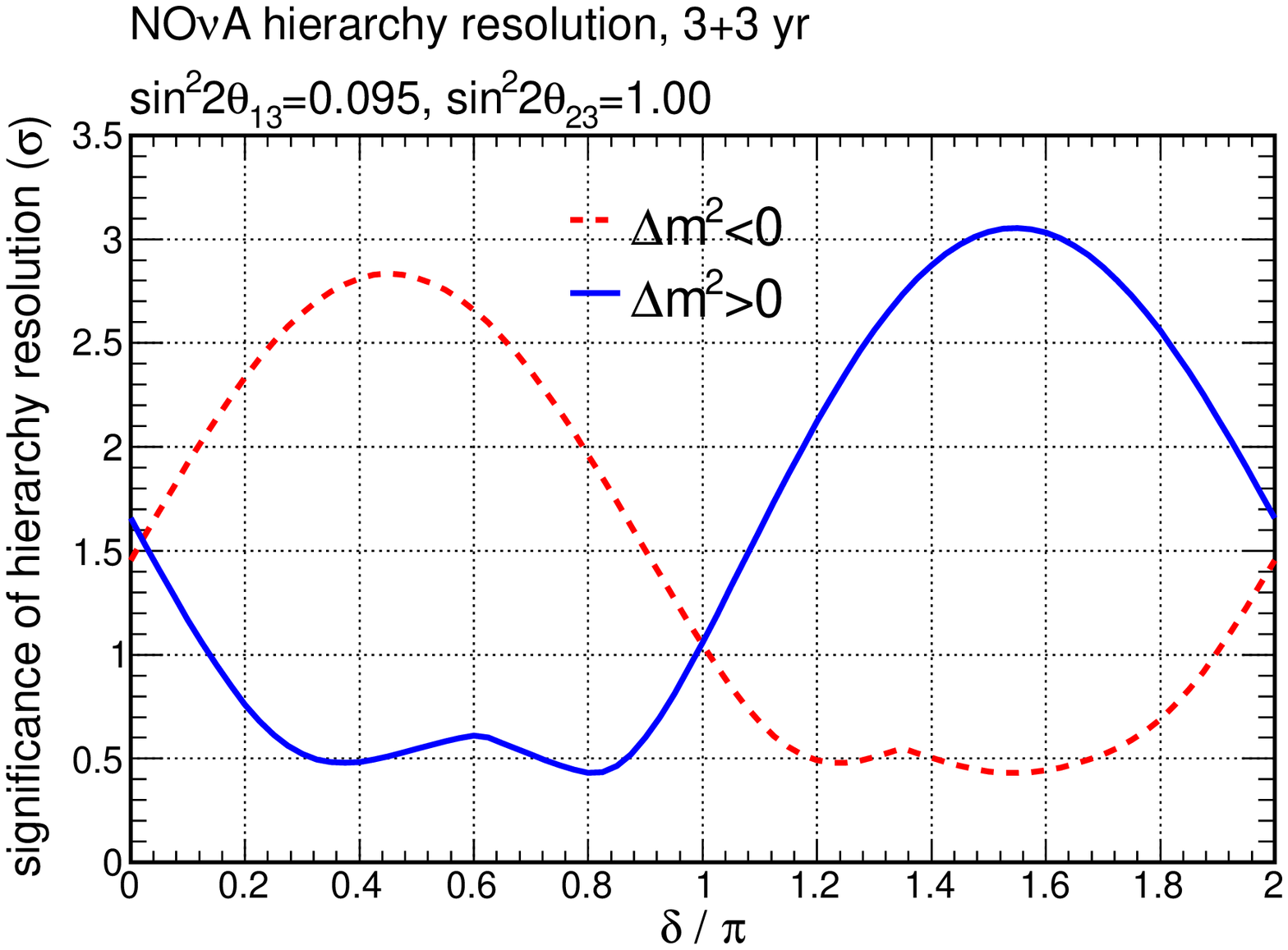}
  \end{minipage}
  \begin{minipage}{.5\linewidth}
    \includegraphics[width=\linewidth]{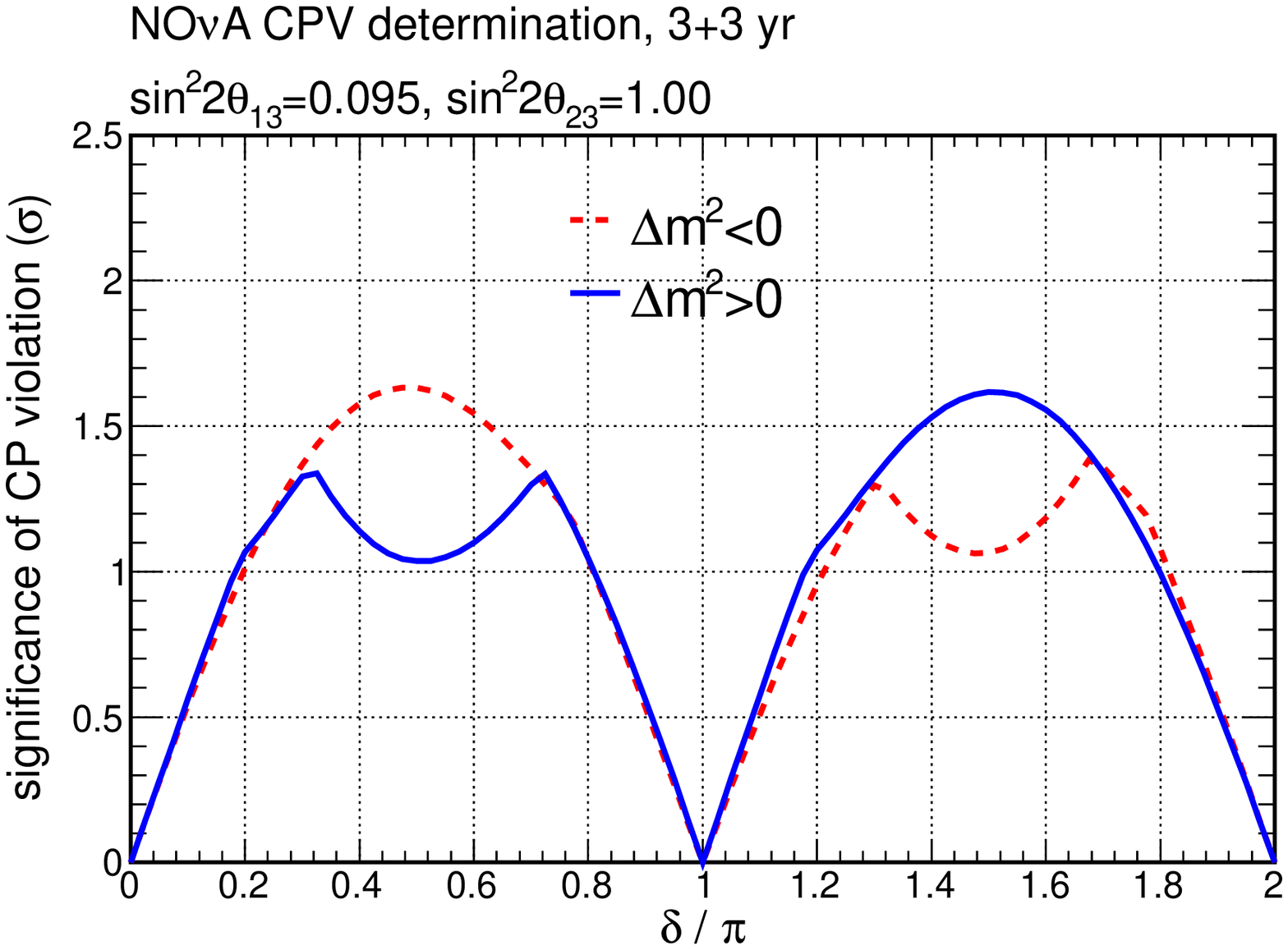}
  \end{minipage}\\
\begin{minipage}{.5\linewidth}
  \includegraphics[width=\linewidth]{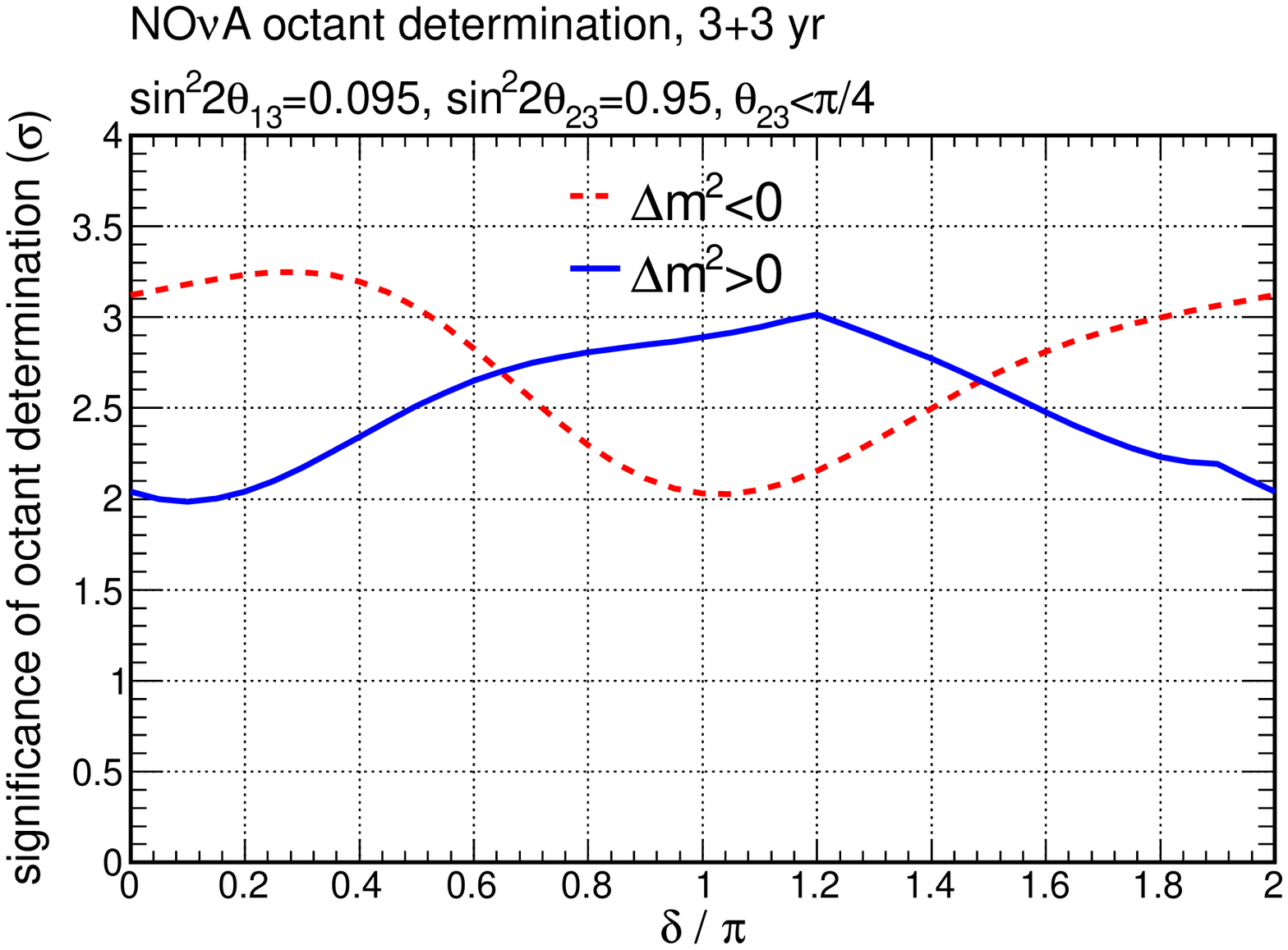}
\end{minipage}
  \begin{minipage}{.5\linewidth}
  \caption{Sensitivity of NO$\nu$A to resolve the mass hierarchy (top left),
    discover $\mathcal{CP}$ violation (above), and resolve the $\theta_{23}$
    octant (left) as a function of the true value of $\delta_{CP}$.
    The solid blue curves are in the case that the true mass hierarchy is
    normal, the dashed red curves are for the inverted case. In each case,
    $\sin^22\theta_{13}=0.095$ is assumed. For hierarchy and $\mathcal{CP}$
    violation $\sin^22\theta_{23}=1$ is assumed, the octant sensitivity assumes
    $\sin^22\theta_{23}=0.95$, first octant.}
  \label{fig:hie_cpv_oct}
  \end{minipage}
\end{figure}

Table \ref{tbl:counts} also shows the representative event counts for the
sample of electron neutrino candidates. As with MINOS, sensitivity to the mass
hierarchy, $\theta_{23}$ octant, and $\delta_{CP}$ comes from combining the
disappearance and appearance analyses, plus constraints on $\theta_{13}$ from
reactor experiments. Figure \ref{fig:hie_cpv_oct} shows the sensitivity of this
analysis to these three parameters, as a function of the true values of
$\delta_{CP}$ and the mass hierarchy. For the mass hierarchy, there are
degenerate regions where it is not possible to disentangle the matter effects,
which give information about the mass hierarchy, from true
$\mathcal{CP}$-violation driven by $\delta_{CP}$. But for favourable
combinations $2-3\sigma$ evidence for the mass hierarchy can be
obtained. Determination of the $\theta_{23}$ octant is much less sensitive to
$\delta_{CP}$ and can be achieved at better than $2\sigma$ for all values of
$\delta_{CP}$ if $\sin^22\theta_{23}=0.95$. The lower octant is assumed in
order to match the best fit from MINOS, performance for the upper octant is
slightly better.
Figure \ref{fig:nova_slices} shows predicted $\Delta\chi^2$ slices as a
function of $\delta_{CP}$, to be compared with the analogous Figure
\ref{fig:minos_slices} from MINOS data. The true parameters are chosen to match
the MINOS best fit (inverted hierarchy, lower octant,
$\delta_{CP}\sim{\pi\over2}$). If the true parameters are indeed in this
region, NO$\nu$A would reject parts of phase space (in particular the
combination of normal hierarchy and upper octant) at very high significance.

\section{Conclusion}

The MINOS experiment has collected a large sample of neutrino interactions in
over six years of running, in both neutrino and antineutrino mode. A combined
fit to this data gives the world's most precise measurement of $|\Delta
m^2_{32}|$, and hints for the neutrino mass hierarchy and $\theta_{23}$
octant. These detectors will continue taking data in the NO$\nu$A-era beam as
MINOS+.

Construction of the NO$\nu$A experiment is progressing well. First neutrinos
have been found in the Far Detector, and projections show good sensitivity to
the mass hierarchy and $\theta_{23}$ octant, in addition to improved
measurements of the atmospheric mixing parameters. Results from the next few
years, in combination with other experiments, should provide significant
information on these questions for the first time.


\section*{References}

\begin{thebibliography}{9}
\bibitem{ref:minos_numu} Adamson P {\it et al.} 2013 {\it Phys. Rev. Lett.} {\bf 110}, 251801 
\bibitem{ref:minos_nue} Adamson P {\it et al.} 2013 {\it Phys. Rev. Lett.} {\bf 110}, 171801 
\bibitem{ref:minos_comb} Adamson P {\it et al.} 2014 {\it Preprint} hep-ex/1403.0867, submitted to {\it Phys. Rev. Lett.}
\bibitem{ref:tdr} Ayres D S {\it et al.} 2007 FERMILAB-DESIGN-2007-01
\bibitem{ref:pdg} Beringer J {\it et al.} 2012 {\it Phys. Rev. D} {\bf 86}, 010001
\end{thebibliography}
\end{document}